\documentclass[runningheads]{llncs}
\usepackage[pdftex]{graphicx}
\begin{document}
\title{Evaluation Framework for Performance Limitation of Autonomous Systems under Sensor Attack}
\titlerunning{Evaluation Framework for Performance Limitation under Sensor Attack}
%
\author{
  Koichi Shimizu\inst{1} \and
  Daisuke Suzuki\inst{1,4} \and
  Ryo Muramatsu\inst{1} \and
  Hisashi Mori\inst{1} \and
  Tomoyuki Nagatsuka\inst{2} \and
  Tsutomu Matsumoto\inst{3,4}
}
\authorrunning{K. Shimizu et al.}
%
\institute{
  Mitsubishi Electric, Kanagawa, Japan \and
  Mitsubishi Electric Engineering, Kanagawa, Japan \and
  Yokohama National University, Kanagawa, Japan \and
  National Institute of Advanced Industrial Science and Technology, Tokyo, Japan
}
\maketitle              
\begin{abstract}

Autonomous systems such as self-driving cars rely on sensors to perceive the surrounding world.
Measures must be taken against attacks on sensors, which have been a hot topic in the last few years.
For that goal one must first evaluate how sensor attacks affect the system,
i.e. which part or whole of the system will fail if some of the built-in sensors are compromised,
or will keep safe, etc.
Among the relevant safety standards, ISO/PAS 21448 addresses the safety of road vehicles
taking into account the performance limitations of sensors, but leaves security aspects out of scope.
On the other hand, ISO/SAE 21434 addresses the security perspective
during the development process of vehicular systems,
but not specific threats such as sensor attacks.
As a result the safety of autonomous systems under sensor attack is yet to be addressed.
%
In this paper we propose a framework that combines safety analysis for scenario identification,
and scenario-based simulation with sensor attack models embedded.
Given an autonomous system model, we identify hazard scenarios caused by sensor attacks,
and evaluate the performance limitations in the scenarios.
We report on a prototype simulator for autonomous vehicles with radar, cameras and LiDAR
along with attack models against the sensors.
Our experiments show that our framework can evaluate how the system safety changes
as parameters of the attacks and the sensors vary.

\end{abstract}
\section{Introduction}

\begin{figure}
  \centering
  \includegraphics[scale=0.4,pagebox=cropbox,clip]{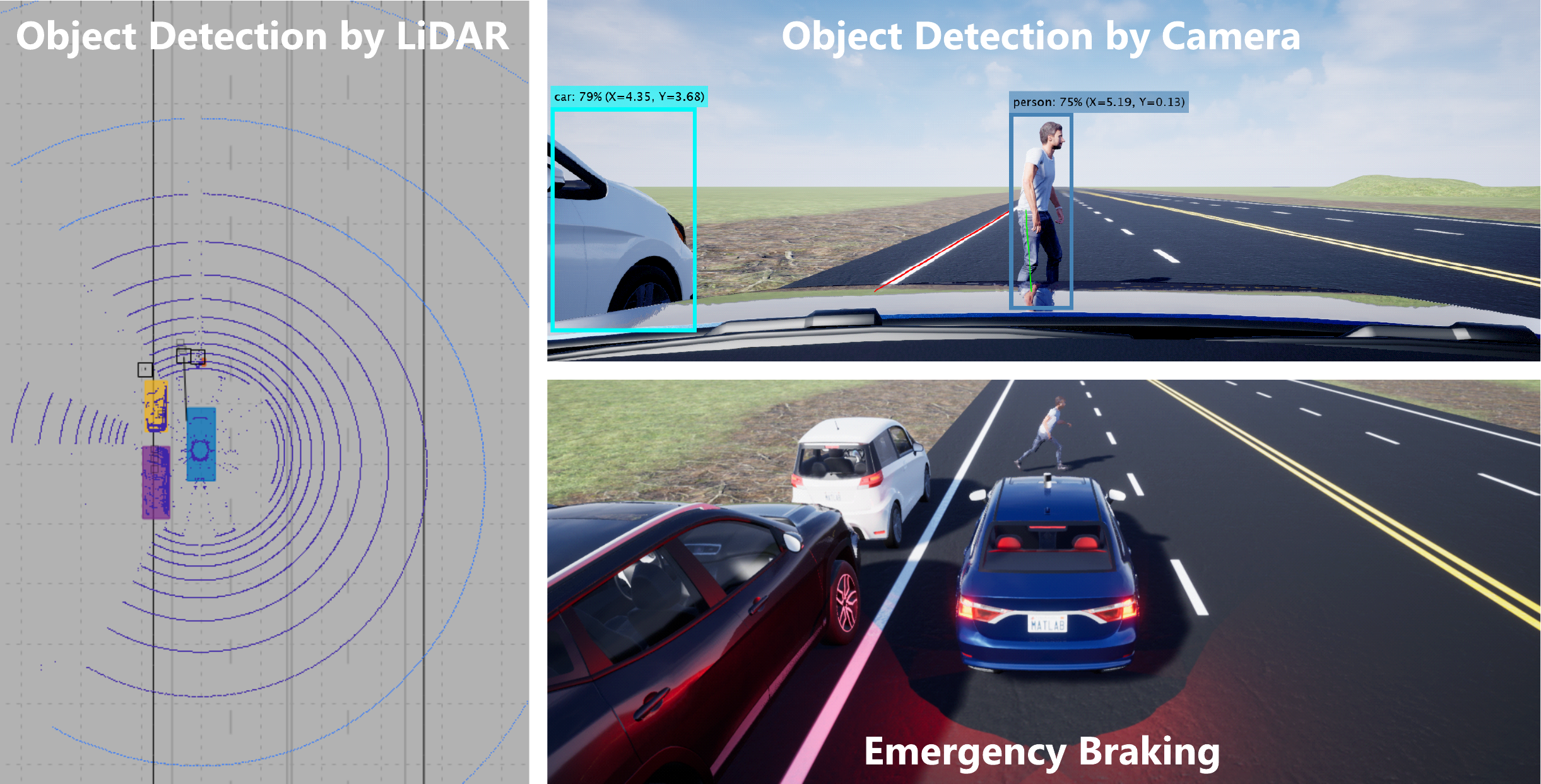}
  \caption{AEB equipped-car with radar, cameras and LiDAR}
  \label{fig:aeb_system}
\end{figure}

Autonomous systems such as autonomous vehicles rely on various sensors
to perceive the surrounding world and decide what to do next.
There have been a lot of reports on attacks against sensors,
e.g. magnetic wheel speed sensors~\cite{10.1007/978-3-642-40349-1_4},
gyro sensors~\cite{190940},
FMCW radar~\cite{Chauhan,10.1145/3338508.3359567}, and LiDAR~\cite{BlindAttack,lidar1},
and against sensor-based autonomous systems~\cite{tibav:36252}.
The safety of autonomous systems against sensor attacks must therefore be assured.
As an illustrative example, we use AEB-equipped cars with radar, cameras and LiDAR
(Fig.~\ref{fig:aeb_system}) throughout the paper.
AEB (Autonomous Emergency Braking) uses the sensors to detect objects around the car,
and measure the distance to and relative speed of the nearest one in front.
If it detects an impending crash, it will dispatch a warning or apply braking.
There is high risk of serious accidents if the sensors are compromised.

To assure the safety of autonomous systems,
scenario-based simulation~\cite{PEGASUS,VeriCAV} is widely accepted as a key tool
because real-world testing for hundreds of millions of miles~\cite{RR-1478-RC} is unrealistic.
One of the issues of scenario-based simulations is how to select a set of relevant scenarios
from the vast space of scenarios consisting of many parameters.
The issue, of course, applies to sensor attack evaluation as well.
In addition, to evaluate the effect of sensor attacks on autonomous systems,
we need autonomous system simulators that embed sensor attack models,
but there has been none thus far.

In this paper we propose a framework to evaluate performance limitations of autonomous systems
in the light of SOTIF.
It combines STAMP/STPA-based safety analysis to identify sensor attack scenarios to be evaluated,
and sensor attack simulation to evaluate the effect of sensor attacks in the scenarios.
We elaborate on safety analysis steps and results for AEB-equipped cars,
and provide a prototype of a sensor attack simulator and examples of evaluation using it.

\paragraph{Contributions}
The main contributions of this paper are threefold:
\begin{itemize}
  \item Evaluation framework of performance limitations that combines safety analysis and
    sensor attack simulation (Section~\ref{sect:eval_framework}).
  \item Method of attack scenario identification based on STAMP/STPA safety analysis
    together with concrete results for AEB (Section~\ref{sect:safety_analysis}).
  \item Autonomous system simulator with sensor attack models embedded,
    and a prototype for AEB together with evaluation examples (Section~\ref{sect:sensor_attack_simulation}).
\end{itemize}

\section{Evaluation Framework Based on SOTIF Process}
\label{sect:eval_framework}

\subsection{Relevant standards and SOTIF}

ISO~26262~\cite{26262} and ISO/PAS~21448~\cite{SOTIF} are safety standards for road vehicles.
The former addresses functional safety as the absence of unreasonable risks caused by failures;
The latter complements functional safety, addressing SOTIF (Safety Of The Intended Functionality)
as the absence of unreasonable risks due to intended functionality or performance limitation.
SOTIF takes into account sensors that advanced functionalities these days rely on.
ISO/SAE~21434~\cite{21434} addresses the security aspects of road vehicles.
It focuses on security risk management during the development process,
and specific attacks are out of scope.

\begin{figure}
  \centering
  \includegraphics[scale=0.4,pagebox=cropbox,clip]{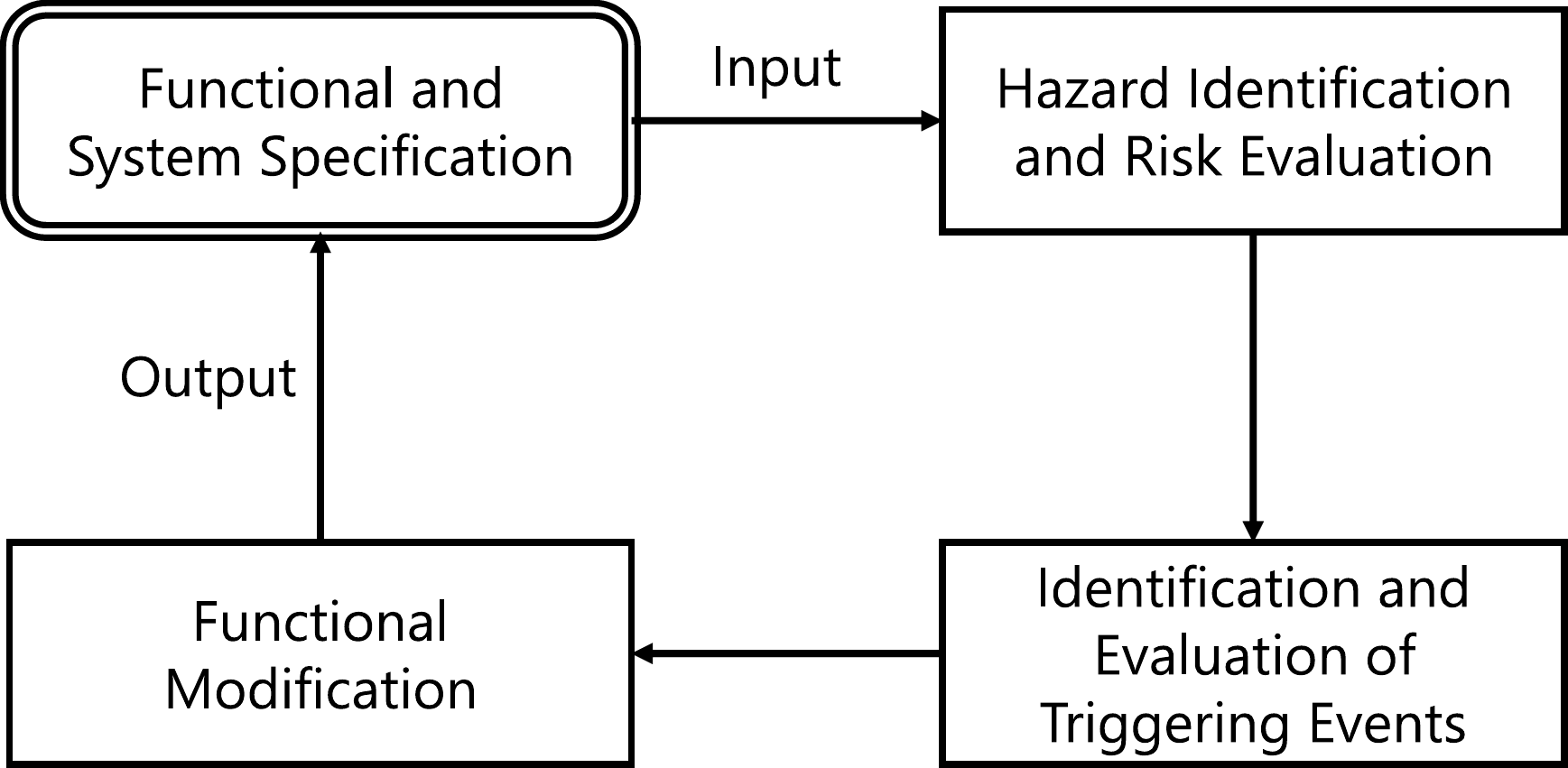}
  \caption{SOTIF improvement process}
  \label{fig:sotif_process}
\end{figure}

The notion of performance limitation in SOTIF with sensors in mind
is compatible with evaluating how sensor attacks affect the system, more specifically,
which part or whole of the system will fail
if some of the built-in sensors are compromised, or will keep safe nevertheless, etc.
We therefore construct an evaluation framework based on the improvement process of SOTIF
(Fig.~\ref{fig:sotif_process}).

Fig.~\ref{fig:sotif_process} depicts a cycle process in which
Functional and System Specification is the starting point,
hazard scenarios are identified for it,
and functions are modified to mitigate the hazard factors.
Performance requirements for sensors are thereby defined at the design stage.
On the other hand, model-based design is widely accepted
for autonomous systems such as vehicles and robots.
It helps evaluate and improve the specification in a continuous manner
from the early stages of development by using an executable specification that can be simulated,
called a model, throughout development.
We adopt a model-based design framework.

\subsection{Evaluation Framework}

\begin{figure}
  \centering
  \includegraphics[scale=0.4,pagebox=cropbox,clip]{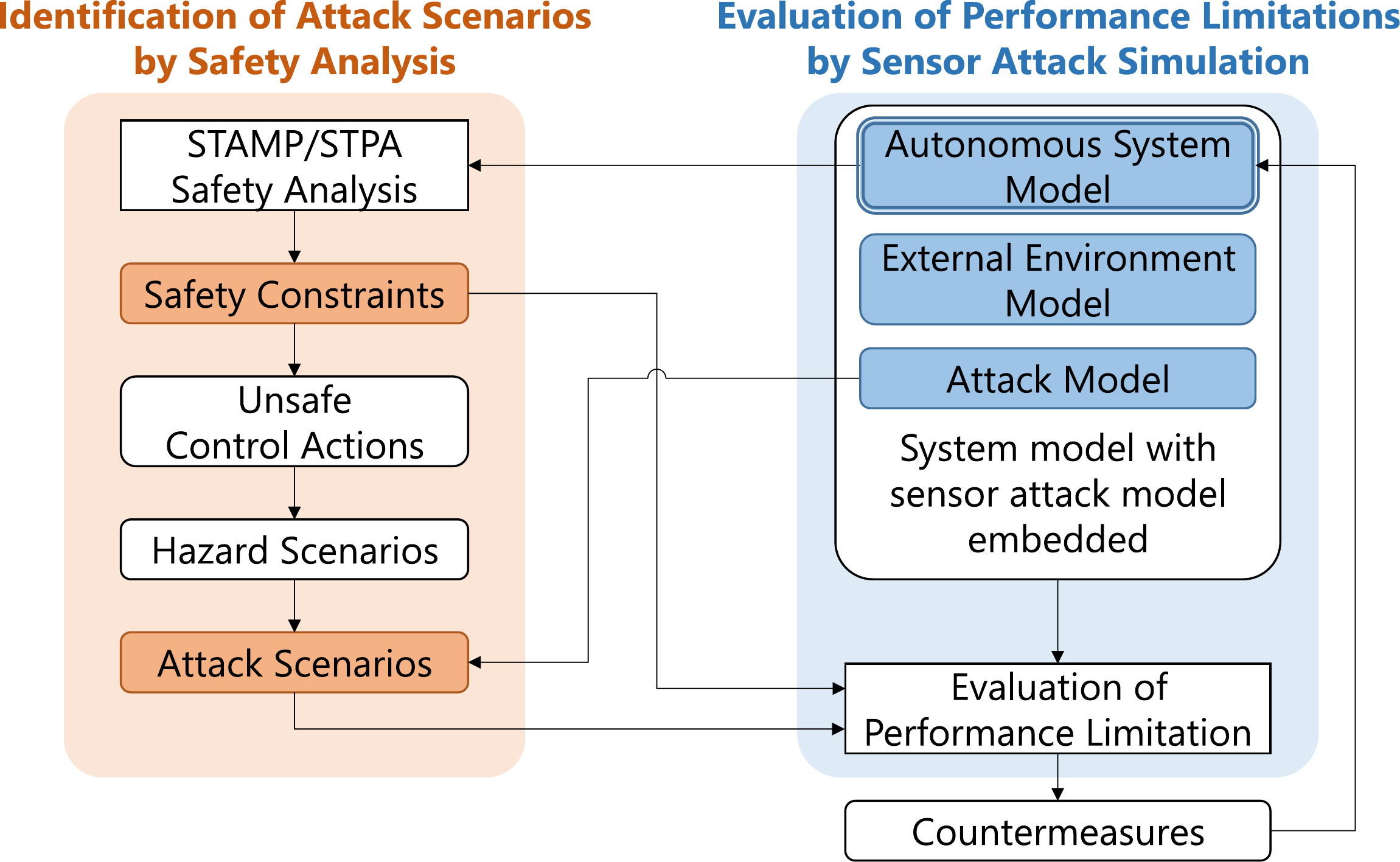}
  \caption{Evaluation framework for performance limitation under sensor attacks}
  \label{fig:eval_framework}
\end{figure}

We present an evaluation framework for performance limitation under sensor attacks
(Fig.~\ref{fig:eval_framework}).
The framework is a combination of STAMP/STPA-based safety analysis
for identifying sensor attack scenarios (Section~\ref{sect:safety_analysis}),
and sensor attack simulation for evaluating the performance limitations in the scenarios
(Section~\ref{sect:sensor_attack_simulation}).
In our framework, the starting point is Autonomous System Model,
the model of the target autonomous system that is created by model-based design.
Given a system model, we extract a control and feedback structure from it
to be analyzed by the left side of the framework.
Once attack scenarios have been identified by the analysis,
we revert the scenarios to the right side of the framework
to be evaluated by sensor attack simulation.

\section{Identifying Attack Scenarios Using STAMP/STPA}
\label{sect:safety_analysis}

\subsection{STAMP/STPA Safety Analysis}

In general, safety analysis is used to identify scenarios that can lead to hazards.
Examples of safety analysis methods include
FTA (Fault Tree Analysis)~\cite{FTA}, FMEA (Failure Mode and Effect Analysis)~\cite{FMEA},
and STAMP/STPA~\cite{STPA}.
While FTA and FMEA focus on hazards caused by component failures,
STAMP/STPA~\footnote[1]{
  STAMP (Systems Theoretic Accident Model and Processes)
  is an accident causality model based on system theory,
  which underpins the analysis method STPA (System-Theoretic Process Analysis).
}
takes the view that hazards can also occur
as a result of unintended interactions between components
even if none of them has any failure.
The view is compatible with SOTIF, and we therefore use STAMP/STPA.

\subsection{Analysis Steps and Results}
\label{sect:stpa_results}

We extract a control and feedback structure to be analyzed from the target system model.
\begin{figure}
  \centering
  \includegraphics[scale=0.4,pagebox=cropbox,clip]{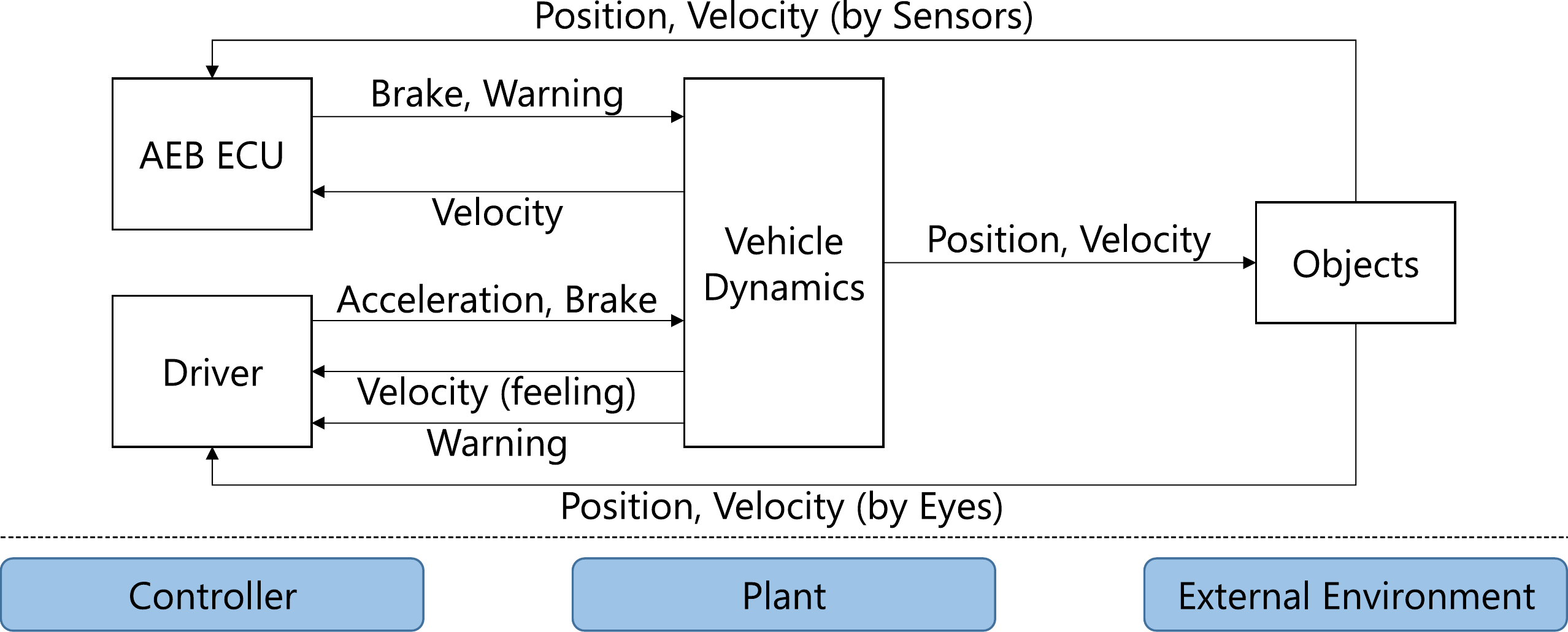}
  \caption{Control and feedback structure of the target AEB-equipped vehicle}
  \label{fig:control_feedback_structure}
\end{figure}
Fig.~\ref{fig:control_feedback_structure} shows the extracted structure,
which consists of the fewest components possible for brevity,
e.g. sensors are not separated from AEB ECU.
The labels at the bottom indicate the correspondence to Fig.~\ref{fig:sensor_attack_simulator}.

\subsubsection{Safety Constraints}

STAMP/STPA first defines hazards
that can lead to losses (e.g. injury, a loss of life, etc.),
and safety constraints as the inverse of hazards.
Safety constraints thereby specify the conditions to be satisfied to keep the system safe,
and we therefore use them as evaluation criteria for performance limitation.
For the current example, we define five safety constraints,
one of which is the following:
\textit{SC1: When the nearest object in front is within a defined distance,
brake must be applied within a defined period of time
to decelerate and stop the ego vehicle}.
We refer to it as \textit{SC1} and use it as an example.

\subsubsection{Unsafe Control Actions}
The next step is to identify UCAs (Unsafe Control Actions) that can break the safety constraints.
STAMP/STPA offers a systematic method for it
by categorizing the causal relationship between control actions and hazards into four types:
1) providing, 2) not providing, 3) too early, too late, and 4) stopped too soon, applied too long.
For the current example, we identify 21 UCAs in total,
14 of which are related to AEB.

\subsubsection{Hazard Scenarios}
The final step of STAMP/STPA is to identify hazard scenarios based on the UCAs.
To create scenarios in a systematic manner, we use hint words supported by
STAMP Workbench~\cite{Workbench},
e.g. 1) Control input or external information wrong or missing,
2) Inadequate or missing feedback, Feedback Delays, etc.
We identify 15 scenarios for the current example.

\subsubsection{Attack Scenarios}
\begin{figure}
  \centering
  \includegraphics[width=\textwidth,pagebox=cropbox,clip]{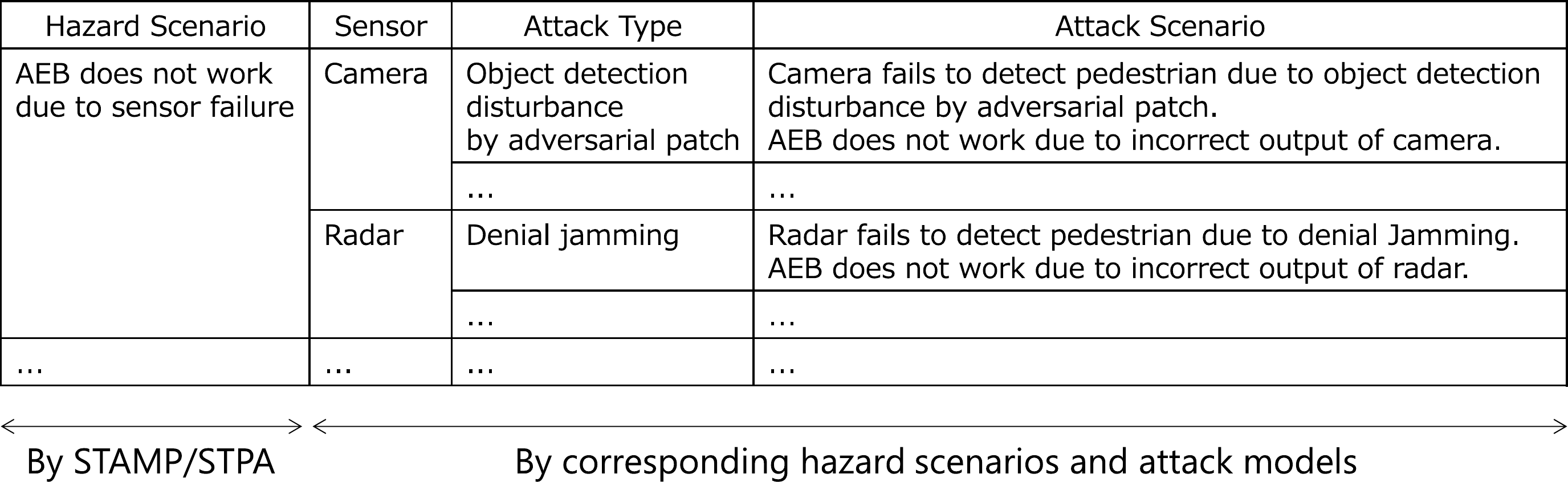}
  \caption{Examples of identified attack scenarios}
  \label{fig:attack_scenarios}
\end{figure}

We create attack scenarios from hazard scenarios
by linking sensor attacks to the causes of hazards.
For that purpose we gather a list of sensor attacks from existing works
that are relevant to AEB-equipped vehicles with radar, cameras and LiDAR.
For radar, we consider denial jamming that prevents object detection~\cite{Tanis,tibav:36252},
and deception jamming on the range~\cite{Chauhan,Tanis,Nashimoto2021,Chen,tibav:36252}
and on the velocity~\cite{Tanis};
For cameras, adversarial patches~\cite{Patch} that disturb the object detection algorithm;
For LiDAR, blinding attacks~\cite{BlindAttack} that perturb the measuring light.
We categorize the attacks into 11 types according to the events that they cause,
and thereby are able to link the attacks to the hazard scenarios.
We identify 102 attack scenarios in total for the current example.
Fig.~\ref{fig:attack_scenarios} shows example scenarios that can break \textit{SC1}.
The scenarios are yet to be concretized for use by the sensor attack simulator;
They are later embedded in operational scenarios
(see Fig.\ref{fig:scenario_cpno}).

\section{Evaluating Performance Limitations under Sensor Attacks}
\label{sect:sensor_attack_simulation}

\begin{figure}
  \centering
  \includegraphics[scale=0.4,pagebox=cropbox,clip]{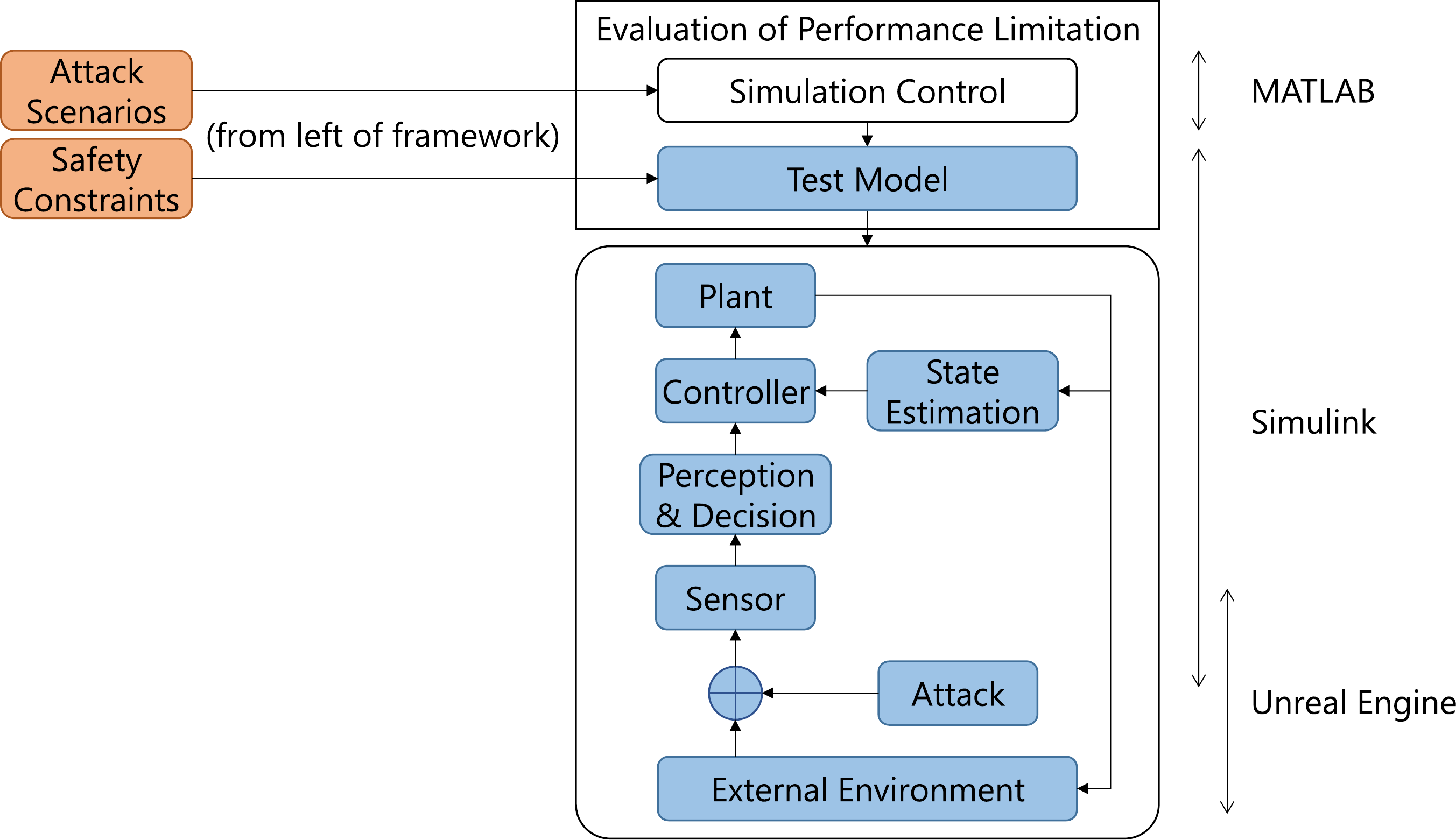}
  \caption{Top-level structure of our sensor attack simulator}
  \label{fig:sensor_attack_simulator}
\end{figure}

We present a sensor attack simulator to realize
the right side of the framework (Fig.~\ref{fig:eval_framework}).
The top-level structure is shown in Fig.~\ref{fig:sensor_attack_simulator}.
Given the safety constraints and attack scenarios,
it evaluates the performance limitations of the target autonomous system
by testing if the system satisfies the safety constraints in the attack scenarios.
We choose MATLAB~\cite{MATLAB}/Simulink~\cite{Simulink} as a platform
widely used in model-based design together with Unreal Engine~\cite{UE}
to implement the external environment and its boundaries with sensors and attacks.

We build a prototype of a sensor attack simulator
for an AEB-equipped car with radar, cameras and LiDAR.

\subsection{Verification of Safety Constraints}
\label{sect:verif_sc}

The simulator must check if the target system satisfies the safety constraints,
and if not, stop running.
There are largely two methods for such evaluation: conventional testing and formal verification.
They have their merits and demerits,
and do not exclude but complement each other~\cite{DBLP:conf/icse/BennionH14,7150771}.
For example, formal verification can give a proof for the verification result
by checking all possible states,
while it can also lead to state explosion as the complexity of a system increases.
One usage is therefore to formally verify the safety-critical part of the system
and to test the system as a whole in a conventional way.
In this paper we use a conventional testing method
with the focus on evaluating the safety of the autonomous system as a whole.

For the current example of AEB,
the safety constraint \textit{SC1} states that the AEB control is correct,
which can be evaluated as follows:
The target model maintains the positions and velocities of objects measured by sensors,
and the AEB control calculated from them.
It also maintains the true values of positions and velocities,
and we can use them to calculate the true AEB control.
By comparing the two AEB controls, we can evaluate if \textit{SC1} is met.

\begin{figure}
  \centering
  \includegraphics[scale=0.4,pagebox=cropbox,clip]{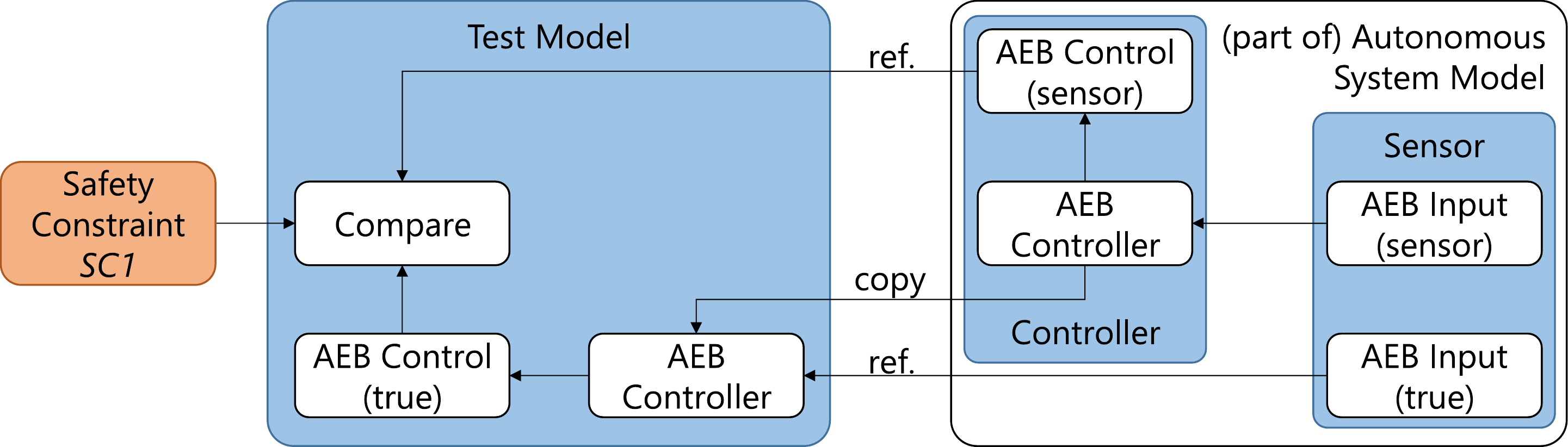}
  \caption{Prototype model for testing the safety constraint \textit{SC1}}
  \label{fig:test_model}
\end{figure}

When we add a model for the evaluation to the simulator,
it is desirable to keep the target system model as unchanged as possible.
Simulink Test~\cite{SimTest} has a mechanism called a test harness
to separate the model for testing from the model under test.
Fig.~\ref{fig:test_model} shows the resultant test model for \textit{SC1} in our prototype.
Implemented as a test harness,
the test model refers to and copies from the target model, but never changes it.

\subsection{Sensor Attack Simulator}
\label{sect:sensor_attack_simulator}

The main body of the sensor attack simulator consists of seven models described below.
All but the Attack model are assumed to be created through the development of the target system.

\subsubsection{Plant, Controller, State Estimation}

Those are the core of a control system.
Our prototype is built around the vehicle dynamics (Plant),
AEB controller (Controller and State Estimation)
and other peripheral models provided by MathWorks.

\subsubsection{Sensor, Perception \& Decision, Attack}
Those are to be designed considering what types of sensor attacks we want to evaluate.
In this paper we model sensor attacks at the same level of abstraction as sensors and external environment
with the view to evaluating attacks on sensors on their own, on sensor fusion,
and on signal processing.

\begin{figure}
  \centering
  \includegraphics[scale=0.4,pagebox=cropbox,clip]{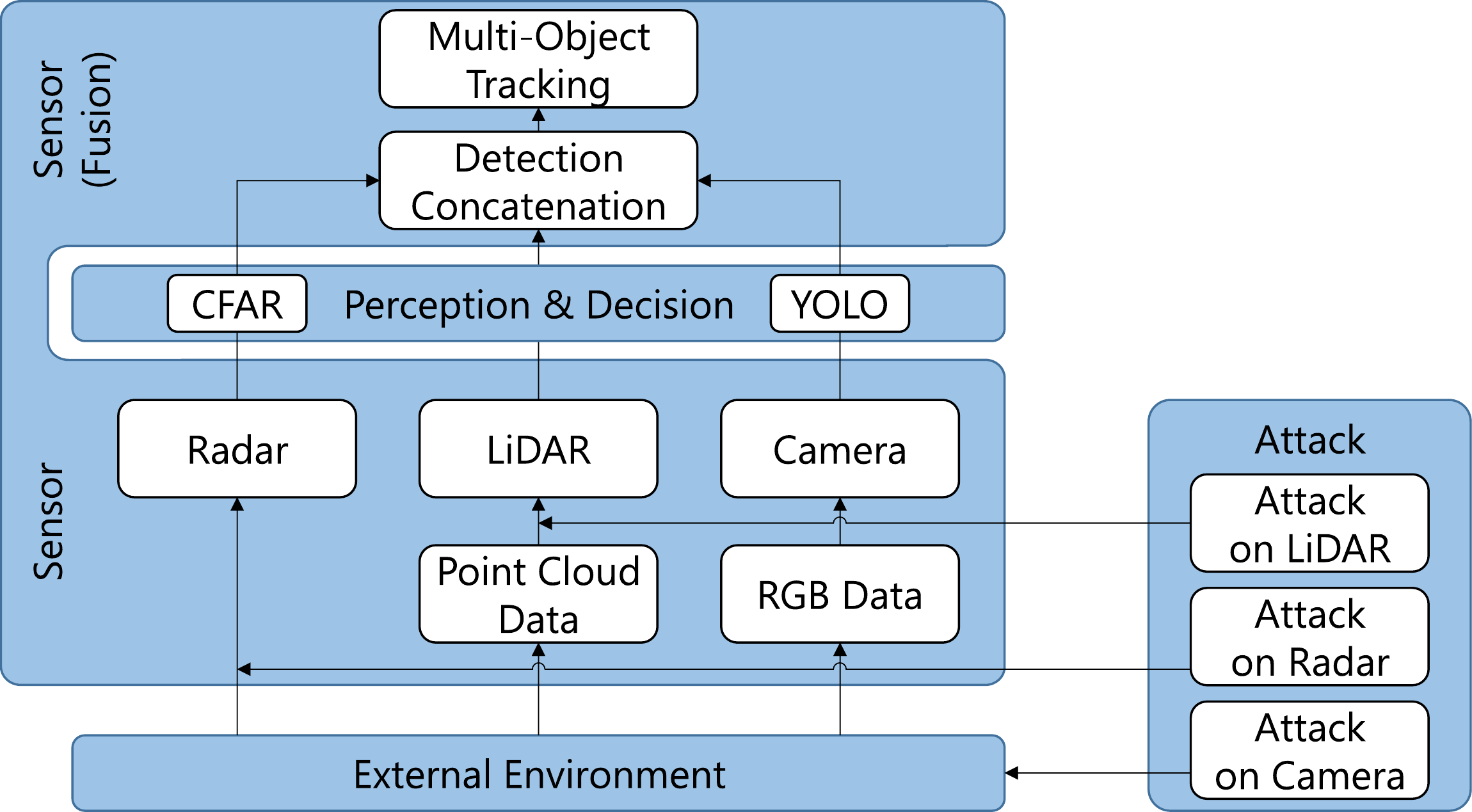}
  \caption{Prototype models of Sensor, Perception \& Decision and Attack}
  \label{fig:sensor_and_attack_models}
\end{figure}

Therefore, the Sensor model is designed to include sensor fusion
as well as separate sensors, namely radar, cameras and LiDAR.
Sensor fusion is further divided into two stages: detection concatenation and multi-object tracking.
The Perception \& Decision model is designed to include object detection algorithms
CFAR (Constant False Alarm Rate) for the radar and YOLO (You Only Look Once) v2~\cite{YOLO} for the cameras.
The Attack model considers those algorithms as well as the sensors on their own.
The resulting models of Sensor, Perception \& Decision
and Attack are shown in Fig.~\ref{fig:sensor_and_attack_models}.

\begin{figure}
  \centering
  \includegraphics[scale=0.4,pagebox=cropbox,clip]{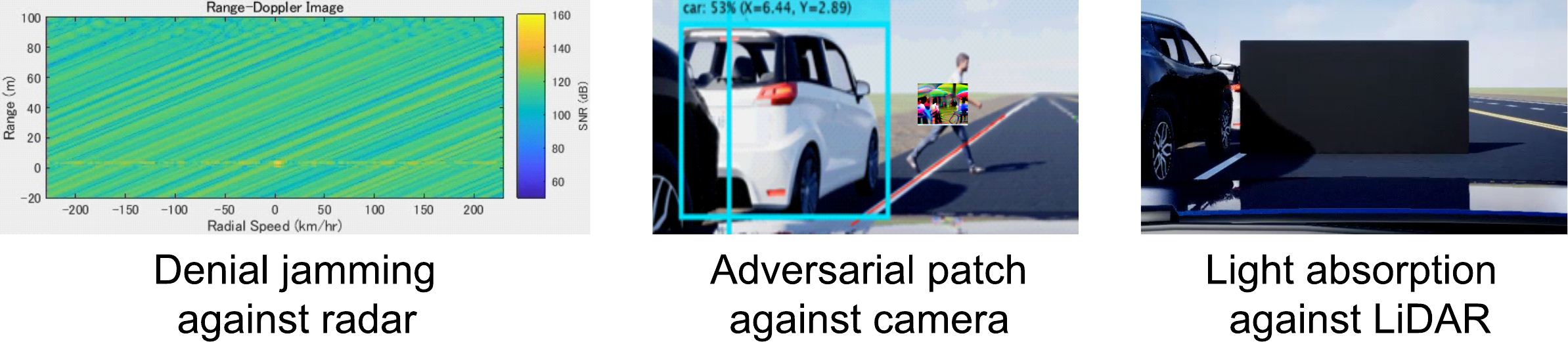}
  \caption{Examples of sensor attack simulation supported by the prototype}
  \label{fig:attack_repertoire}
\end{figure}

Our collection of attack models are as described in Section~\ref{sect:stpa_results};
We model the gathered list of sensor attacks that are relevant to
AEB-equipped vehicles with radar, cameras and LiDAR.
We show examples of sensor attack simulation supported by the prototype
in Fig.~\ref{fig:attack_repertoire}.

\subsubsection{External Environment}

It models the external environment surrounding the target system,
e.g. nearby objects, how they are perceived by the sensors,
and the positional relationship between the target system and the other objects.
It also defines the temporal development of the target system and the environment
as operational scenarios.

\begin{figure}
  \centering
  \includegraphics[scale=0.4,pagebox=cropbox,clip]{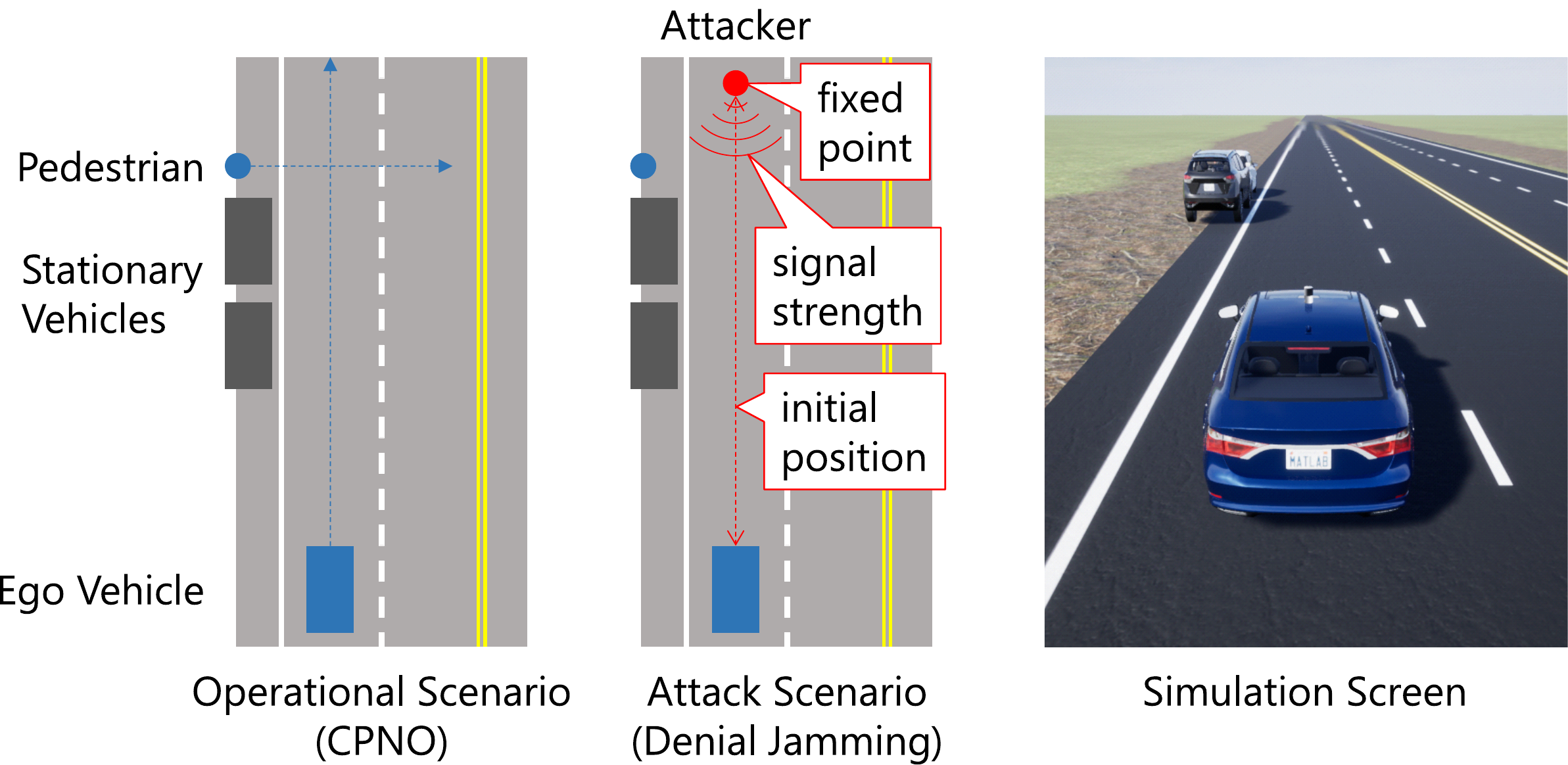}
  \caption{CPNO scenario and an example of denial jamming in CPNO}
  \label{fig:scenario_cpno}
\end{figure}

Our prototype models the external environment that comply with
the evaluation criteria of AEB in JNCAP~\cite{JNCAP} and Euro NCAP~\cite{EuroNCAP},
and supports the complete set of operational scenarios:
largely, five scenarios of car detection and 11 scenarios of pedestrian and cyclist detection,
and a total of 278 scenarios with parameter variations.
As an example, Fig.~\ref{fig:scenario_cpno} shows the CPNO (Car-to-Pedestrian Nearside Obstructed)
scenario in JNCAP, where
the ego vehicle travels forward towards a pedestrian crossing its path from the nearside
who is out of sight at first due to stationary vehicles in between.

Fig.~\ref{fig:scenario_cpno} also shows an example attack scenario embedded in CPNO,
in which denial jamming is applied from a fixed point in front
with the attacker's initial position and signal strength variable.
Those attack settings are not specified in Fig.~\ref{fig:attack_scenarios}.
Only when we determine the operation scenario and embed an attack scenario into it,
can we concretize the attack settings.

\subsection{Evaluation Examples Using the Prototype}
\label{sect:eval_attack_parameters}

We show three examples of evaluation using the prototype:
one about attack parameters, and two about sensor design parameters.
We use the attack scenario of denial jamming in CPNO (Fig.~\ref{fig:scenario_cpno}).
For the sake of brevity, the simulation stops when the car crashes into the pedestrian
instead of when the safety constraints such as \textit{SC1} are not met.

\subsubsection{Jamming Attack on the Radar}

We evaluate the effect of jamming attack on the radar
with respect to two parameters: the attacker's position and signal strength.
The other parameters are fixed: the velocity of the ego vehicle is 25 [km/h],
and the signal strength of the ego vehicle, 10 [dBm].
The cameras and LiDAR are not used for AEB control for the sake of evaluating the radar alone.

\begin{figure}
  \centering
  \includegraphics[scale=0.4,pagebox=cropbox,clip]{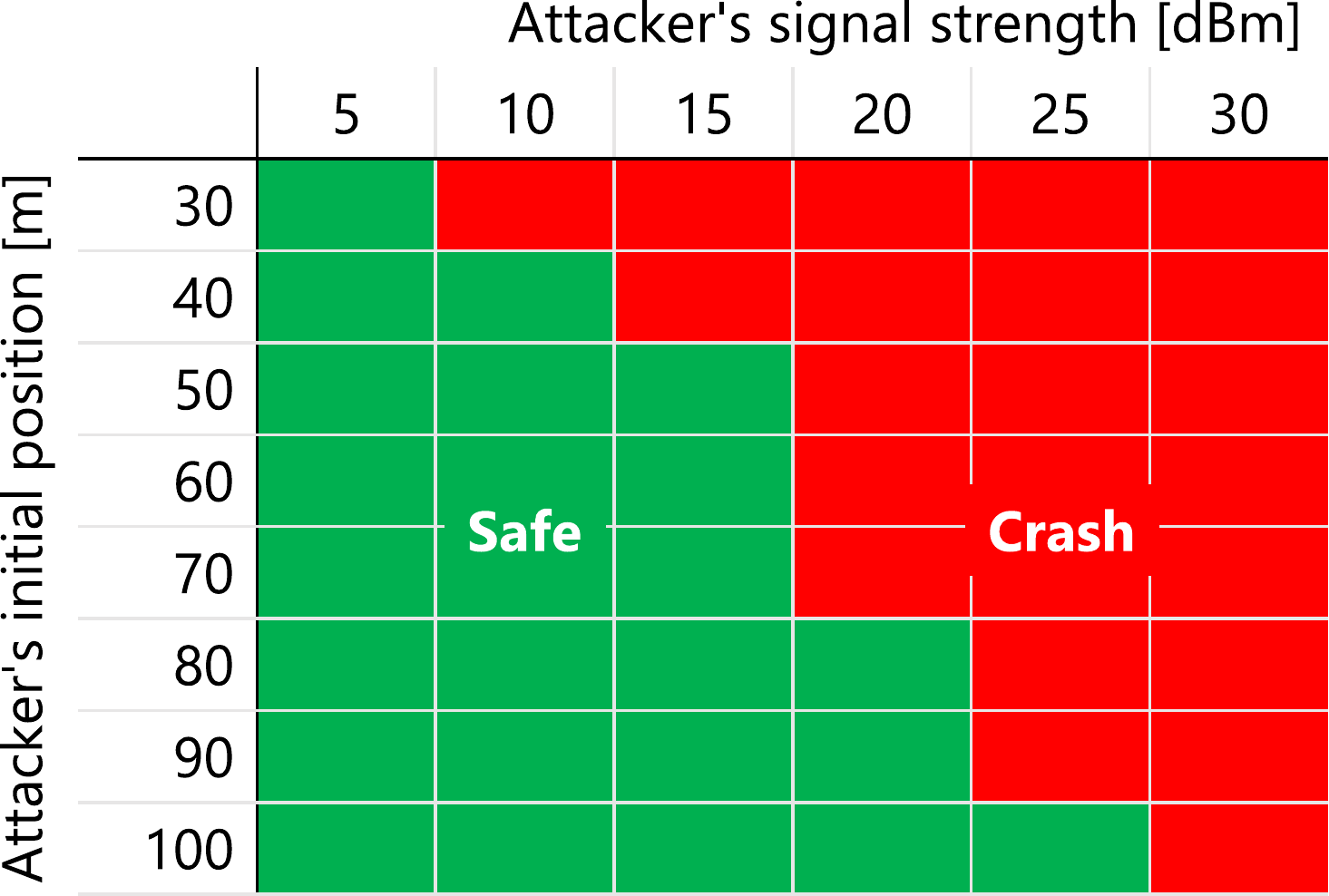}
  \caption{Evaluation result with respect to radar jamming parameters.
  The detections by camera and LiDAR are not used for AEB control.}
  \label{fig:result_attack_parameters}
\end{figure}

Fig~\ref{fig:result_attack_parameters} shows the results.
Each element of the matrix denotes whether the car crashes into the pedestrian (Crash)
or not (Safe).
The result is as expected: the stronger the attacker's signal is,
or the nearer the attacker's position is, the more likely the attack is to succeed.
That proves the validity of the simulation.

\subsubsection{Detection Concatenation}

As an example of sensor fusion,
we evaluate the effect of concatenation of the radar and camera.
The attacker's position and signal strength are set to 30 [m] and 10 [dBm],
and all the other conditions are the same.
The LiDAR is not used for AEB control.

\begin{figure}
  \centering
  \includegraphics[scale=0.4,pagebox=cropbox,clip]{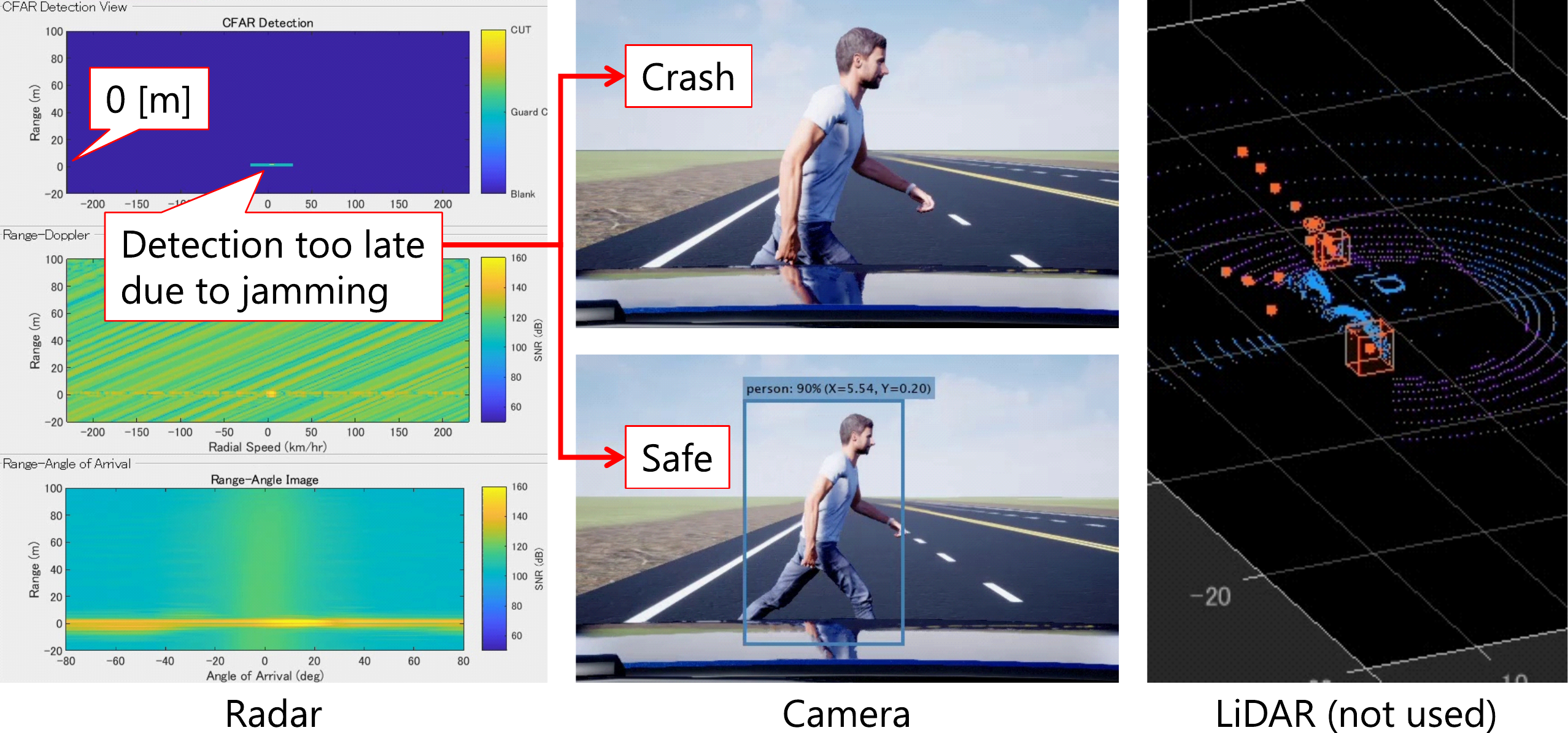}
  \caption{Evaluation result for Detection Concatenation.
  The detection by LiDAR is not used for AEB control.}
  \label{fig:result_concatenation}
\end{figure}

Fig.~\ref{fig:result_concatenation} shows the results.
The leftmost part shows object detection by the radar,
and the central part, footage of the front camera of the ego vehicle,
in which the upper half is in the case of the radar alone,
and the lower half, the concatenation of the radar and camera.
Due to the jamming, it is only when the distance is close to 0 [m]
that the radar detects the person in front;
It is too late to avoid a crash with the radar alone,
while the car is safely stopped with the concatenation thanks to detection by the camera.
The comparison proves the effectiveness of the concatenation.

\subsubsection{Multi-Object Tracking}
As a second example of sensor fusion,
we evaluate the multi-object tracking algorithm.
Simply put, the algorithm tracks objects by maintaining a list of
object detections by multiple sensors.
To exclude the effect of misdetections, the algorithm confirms the detection
if the same object is detected at least M times out of N sensing periods.
Therefore, the greater the ratio M/N is, the more accurate the detection becomes.
If we increase N with a fixed ratio M/N,
we expect to eliminate the effect of variance and further improve the accuracy,
while the algorithm can become more susceptible to attacks
due to the increased processing time.

\begin{figure}
  \centering
  \includegraphics[width=\textwidth,pagebox=cropbox,clip]{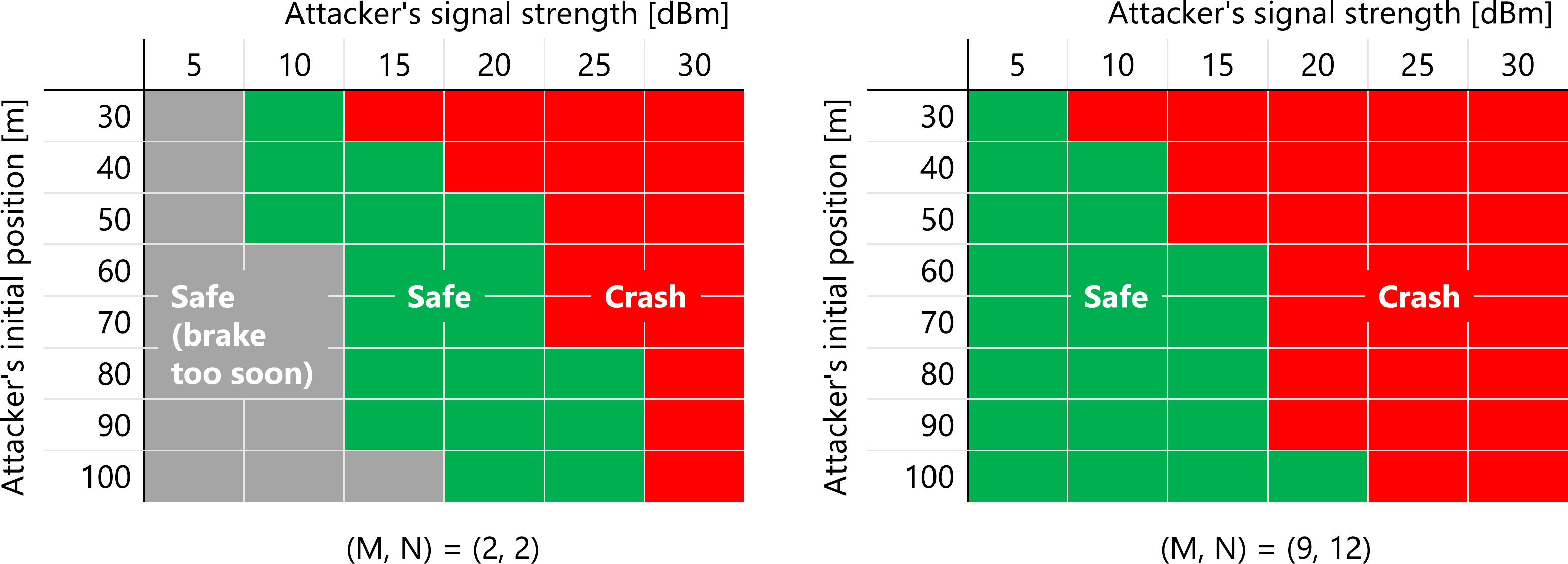}
  \caption{Evaluation results for Multi-Object Tracking with different sets of parameters}
  \label{fig:result_multi_object_tracking}
\end{figure}

We evaluate the effect of the design parameters M and N
in the same attack settings as in Fig.~\ref{fig:result_attack_parameters}.
Fig.~\ref{fig:result_multi_object_tracking} shows the evaluation results
for (M, N) = (2, 2) and (9, 12).
Overall, (M, N) = (2, 2) is safer than (M, N) = (9, 12)
because there are fewer crashes in the former case.
However, there are also cases where the brake is applied too soon,
which can lead to an uncomfortable driving experience.
We can evaluate this kind of trade-off with our simulator.

\section{Related Work}

Coverage by scenario-based simulation for autonomous systems has been
extensively studied~\cite{PEGASUS,VeriCAV,Saigol,Norden,Abdessalem,9176839}.
%
Coverage criteria and techniques to maximize it are investigated in~\cite{9176839}.
Coverage maximizing techniques include automated generation of test scenarios
by random numbers~\cite{Saigol,Norden} and by search algorithms~\cite{Abdessalem}.
In~\cite{Abdessalem}, the authors consider critical test scenarios leading to failures,
which looks suitable for performance limitation evaluation.
For specific systems like AEB,
we have some prior knowledge about scenarios.
In~\cite{PEGASUS}, scenarios are defined in a systematic manner
with six layers such as road, moving objects and environmental conditions.
%
There are other challenges in scenario-based simulation~\cite{VeriCAV}.
Interface between various autonomous system models and simulation tools
is a key issue considering autonomous systems becoming more and more complex.

There have been works on safety analysis of autonomous systems
using STAMP/STPA~\cite{ABDULKHALEQ20152,DAKWAT2018130,Ishimatsu}.
While we use conventional testing in this paper,
formal methods are promising for verifying the safety constraints~\cite{ABDULKHALEQ20152,DAKWAT2018130}.
In \cite{ABDULKHALEQ20152}, the authors embed the safety constraints
to the target model, and formally verify them.
A comparison with FTA is detailed in \cite{Ishimatsu},
which concludes that STAMP/STPA can identify more scenarios.

\section{Conclusion}

We present a framework to evaluate performance limitations of autonomous systems under sensor attacks.
Using a prototype simulator of an AEB-equipped car with radar, cameras and LiDAR,
we show that the framework can identify sensor attack scenarios to be assessed,
and evaluate how attacks on the sensors affect the system safety.

Interface between different models and simulators is a key issue
in evaluation of autonomous systems,
especially when it comes to highly autonomous vehicles becoming more and more complex.
We therefore leave it as future work to modularize
the sensor attack models, e.g. as FMU (Functional Mock-up Unit),
to be used in combination with other simulators.
In addition to self-driving cars,
there are a diverse range of critical devices and systems that depend on measurement,
such as robotic systems, medical devices and control systems.
We therefore want to extend our framework to address
attacks and countermeasures about measurement interfaces in general:
what is called \textit{instrumentation security}.

\subsubsection{Acknowledgment.}
This work is partially based on results obtained from the project (JPNP16007)
commissioned by the New Energy and Industrial Technology Development Organization (NEDO).

%
%
%
\bibliographystyle{splncs04}
\bibliography{biblio}

\begin{thebibliography}{10}
\providecommand{\url}[1]{\texttt{#1}}
\providecommand{\urlprefix}{URL }
\providecommand{\doi}[1]{https://doi.org/#1}

\bibitem{EuroNCAP}
{Euro NCAP}, \url{https://www.euroncap.com/en}

\bibitem{JNCAP}
{JNCAP}, \url{https://www.nasva.go.jp/mamoru/en/}

\bibitem{ABDULKHALEQ20152}
Abdulkhaleq, A., Wagner, S., Leveson, N.: A comprehensive safety engineering
  approach for software-intensive systems based on stpa. Procedia Engineering
  \textbf{128},  2--11 (2015), proceedings of the 3rd European STAMP Workshop
  5-6 October 2015, Amsterdam

\bibitem{Abdessalem}
{Ben Abdessalem}, R., {Nejati}, S., {C. Briand}, L., {Stifter}, T.: Testing
  vision-based control systems using learnable evolutionary algorithms. In:
  2018 IEEE/ACM 40th International Conference on Software Engineering (ICSE).
  pp. 1016--1026 (2018)

\bibitem{DBLP:conf/icse/BennionH14}
Bennion, M., Habli, I.: A candid industrial evaluation of formal software
  verification using model checking. In: Jalote, P., Briand, L.C., van~der
  Hoek, A. (eds.) 36th International Conference on Software Engineering, {ICSE}
  '14, Companion Proceedings, Hyderabad, India, May 31 - June 07, 2014. pp.
  175--184. {ACM} (2014)

\bibitem{Chauhan}
Chauhan, R.: A platform for false data injection in frequency modulated
  continuous wave radar all graduate theses and dissertations. 3964.
  \url{https://digitalcommons.usu.edu/etd/3964} (2014)

\bibitem{Chen}
Chen, H.R.: {FMCW} radar jamming techniques and analysis (2013),
  \url{https://calhoun.nps.edu/handle/10945/37597}

\bibitem{DAKWAT2018130}
Dakwat, A.L., Villani, E.: System safety assessment based on stpa and model
  checking. Safety Science  \textbf{109},  130--143 (2018)

\bibitem{UE}
{Epic Games}: {Unreal Engine}. \url{https://www.unrealengine.com/}

\bibitem{FTA}
IEC: Fault tree analysis ({FTA}). Standard IEC 61025:2006 (2006)

\bibitem{FMEA}
IEC: Failure modes and effects analysis ({FMEA} and {FMECA}). Standard IEC
  60812:2018 (2018)

\bibitem{Workbench}
{IPA}: {STAMP Workbench}.
  \url{https://www.ipa.go.jp/sec/stamp\_wb/manual/index.html}

\bibitem{Ishimatsu}
Ishimatsu, T., Leveson, N.G., Thomas, J., Katahira, M., Miyamoto, Y., Nakao,
  H.: Modeling and hazard analysis using stpa. International Association for
  the Advancement of Space Safety (IAASS) (2010)

\bibitem{26262}
ISO: Road vehicles -- functional safety. Standard ISO 26262:2018 (2018)

\bibitem{SOTIF}
ISO: Road vehicles -- safety of the intended functionality. Standard ISO/PAS
  21448:2019(E) (2019)

\bibitem{21434}
ISO: Road vehicles -- cybersecurity engineering. Standard ISO/SAE DIS
  21434:2020(E) (2020)

\bibitem{7150771}
{Jeppu}, N.Y., {Jeppu}, Y., {Murthy}, N.: Arguing formally about flight control
  laws. In: 2015 International Conference on Industrial Instrumentation and
  Control (ICIC). pp. 378--383 (2015)

\bibitem{RR-1478-RC}
Kalra, N., Paddock, S.M.: Driving to Safety: How Many Miles of Driving Would It
  Take to Demonstrate Autonomous Vehicle Reliability? RAND Corporation, Santa
  Monica, CA (2016)

\bibitem{VeriCAV}
Levermore, T., Peters, A.: {Test framework and key challenges for virtual
  verification of automated vehicles: the VeriCAV project}. In: {39th
  International Conference on Computer Safety, reliability and Security
  (SAFECOMP), Position Paper}. Lisbon, Portugal (Sep 2020),
  \url{https://hal.archives-ouvertes.fr/hal-02931723}

\bibitem{STPA}
Leveson, N., Thomas, J.: {STPA} handbook (2018),
  \url{http://psas.scripts.mit.edu/home/get\_file.php?name=STPA\_handbook.pdf}

\bibitem{tibav:36252}
Liu, J., Yan, C., Xu, W.: Can you trust autonomous vehicles: Contactless
  attacks against sensors of self-driving vehicles. DEF CON (2016),
  https://doi.org/10.5446/36252

\bibitem{MATLAB}
MathWorks: {MATLAB}. \url{https://jp.mathworks.com/products/matlab.html}

\bibitem{Simulink}
MathWorks: Simulink. \url{https://jp.mathworks.com/products/simulink.html}

\bibitem{SimTest}
MathWorks: {Simulink Test}.
  \url{https://jp.mathworks.com/products/simulink-test.html}

\bibitem{10.1145/3338508.3359567}
Miura, N., Machida, T., Matsuda, K., Nagata, M., Nashimoto, S., Suzuki, D.: A
  low-cost replica-based distance-spoofing attack on mmwave fmcw radar. In:
  Proceedings of the 3rd ACM Workshop on Attacks and Solutions in Hardware
  Security Workshop. p. 95–100. ASHES'19, Association for Computing
  Machinery, New York, NY, USA (2019)

\bibitem{Nashimoto2021}
Nashimoto, S., Suzuki, D., Miura, N., Machida, T., Matsuda, K., Nagata, M.:
  Low-cost distance-spoofing attack on fmcw radar and its feasibility study on
  countermeasure. Journal of Cryptographic Engineering  (Jan 2021)

\bibitem{Norden}
Norden, J., O'Kelly, M., Sinha, A.: Efficient black-box assessment of
  autonomous vehicle safety. CoRR  \textbf{abs/1912.03618} (2019),
  \url{http://arxiv.org/abs/1912.03618}

\bibitem{lidar1}
Petit, J., Stottelaar, B., Feiri, M., Kargl, F.: Remote attacks on automated
  vehicles sensors: Experiments on camera and lidar. In: Black Hat Europe
  (2015),
  \url{https://www.blackhat.com/docs/eu-15/materials/eu-15-Petit-Self-Driving-And-Connected-Cars-Fooling-Sensors-And-Tracking-Drivers-wp1.pdf}

\bibitem{YOLO}
{Redmon}, J., {Farhadi}, A.: Yolo9000: Better, faster, stronger. In: 2017 IEEE
  Conference on Computer Vision and Pattern Recognition (CVPR). pp. 6517--6525
  (2017)

\bibitem{Saigol}
Saigol, Z., Peters, A.: Verifying automated driving systems in simulation:
  framework and challenges. In: 25th ITS World Congress (2018)

\bibitem{BlindAttack}
Shin, H., Kim, D., Kwon, Y., Kim, Y.: Illusion and dazzle: Adversarial optical
  channel exploits against lidars for automotive applications. In: Fischer, W.,
  Homma, N. (eds.) Cryptographic Hardware and Embedded Systems -- CHES 2017.
  pp. 445--467. Springer International Publishing, Cham (2017)

\bibitem{10.1007/978-3-642-40349-1_4}
Shoukry, Y., Martin, P., Tabuada, P., Srivastava, M.: Non-invasive spoofing
  attacks for anti-lock braking systems. In: Bertoni, G., Coron, J.S. (eds.)
  Cryptographic Hardware and Embedded Systems - CHES 2013. pp. 55--72. Springer
  Berlin Heidelberg, Berlin, Heidelberg (2013)

\bibitem{190940}
Son, Y., Shin, H., Kim, D., Park, Y., Noh, J., Choi, K., Choi, J., Kim, Y.:
  Rocking drones with intentional sound noise on gyroscopic sensors. In: 24th
  {USENIX} Security Symposium ({USENIX} Security 15). pp. 881--896. {USENIX}
  Association (2015)

\bibitem{9176839}
{Tahir}, Z., {Alexander}, R.: Coverage based testing for v \& v and safety
  assurance of self-driving autonomous vehicles: A systematic literature
  review. In: 2020 IEEE International Conference On Artificial Intelligence
  Testing (AITest). pp. 23--30 (2020)

\bibitem{Tanis}
Tanis, S.: Automotive radar sensors and congested radio spectrum: An urban
  electronic battlefield? In: Analog Dialogue. vol. 52-07 (2018)

\bibitem{Patch}
{Thys}, S., {Ranst}, W.V., {Goedemé}, T.: Fooling automated surveillance
  cameras: Adversarial patches to attack person detection. In: 2019 IEEE/CVF
  Conference on Computer Vision and Pattern Recognition Workshops (CVPRW). pp.
  49--55 (2019)

\bibitem{PEGASUS}
Weber, H., Bock, J., Klimke, J., Roesener, C., Hiller, J., Krajewski, R.,
  Zlocki, A., Eckstein, L.: A framework for definition of logical scenarios for
  safety assurance of automated driving. Traffic Injury Prevention
  \textbf{20}(sup1),  S65--S70 (2019), pMID: 31381437

\end{thebibliography}

\end{document}